\documentclass{article}
\begin{document}
\title{\bf A simple mathematical model for anomalous diffusion via Fisher's information theory}
\author{Marcelo R. Ubriaco\thanks{Electronic address:ubriaco@ltp.upr.clu.edu}}
\date{Laboratory of Theoretical Physics\\Department of Physics\\University of Puerto Rico\\R\'{\i}o Piedras Campus\\
San Juan\\PR 00931, USA}
\maketitle
\begin{abstract}
Starting with the relative entropy based on a previously proposed  entropy function $S_q[p]=\int dx\; p(x)(-\ln p(x))^q$, we find the corresponding
Fisher's information measure. After function redefinition we then maximize
 the Fisher information measure with
respect to the new function and obtain a differential operator that
reduces to a space coordinate second derivative in the $q\rightarrow 1$ limit. We then
propose a simple differential equation for anomalous diffusion and show that its
solutions are a generalization of the functions in the Barenblatt-Pattle solution.
We find that the mean squared displacement, up to a $q$-dependent constant, has a time
dependence according to $<x^2>\sim K^{1/q}t^{1/q}$, where the parameter $q$ takes values  $q=\frac{2n-1}{2n+1}$ (superdiffusion)
and
$q=\frac{2n+1}{2n-1}$ (subdiffusion),  $\forall n\geq 1$.

\end{abstract}
\vspace{0.2in}
PACS number(s): 89.70.Cf, 05.20.-y

\section{Introduction}
During the past two decades the Fisher information measure \cite{Fisher1} has become,
due mainly  to the work of Frieden and coworkers, to play a
fundamental role from where the well known covariant field equations of physics can be extracted \cite{F1}-\cite{F3}. Although the Fisher information measure is
obtained from the relative entropy, it plays a different
role than the Shannon's information measure. As physics is concerned, in statistical mechanics maximization of Shannon's entropy under some constraints gives us the probability
distribution function and therefore  the partition function for a particular system, while in the Fisher information measure
after defining a new function $\phi(p)$ in terms of the probability amplitudes $p(x)$, one obtains
a 'kinetic'  term in terms of $\phi$ leading to the interpretation of the Fisher information
as a kind of lagrangian. In particular,  in Refs. \cite{F1}-\cite{F3} it has been shown
that the familiar relativistic wave equations are obtained maximizing the
expressions $I-J$, where $I$ is the Fisher measure in term of the new functions and $J$
is obtained from the Fourier transform of $I$ and use of the energy constraint. Curiously,
this approach works very well for the case of covariant formulations. For the nonconvariant
case, as it is the case of the Schr\"{o}dinger  and the diffusion equations, the Fisher measure
defines the space coordinates differential operator and thus the time part has to be introduced
by using dimensional analysis. 

This paper is organized as follows. In Section  (\ref{Fisher})
we   review the relation of the Fisher measure and the Kullback entropy,
and briefly explain how the definition of the new function $\phi$ as a function of $p(x)$ leads to
the interpretation of $I$ as a lagrangian. In Section (\ref{relat}) we introduce the
relative entropy for our entropy function and we obtain, following the method in Ref \cite{Vs}),
the corresponding Fisher measure, and then after function redefinition we find
coordinate differential operator. In this section we also  obtain the eigenfunctions of this
differential operator. In Section (\ref{anomalous}) we show that this formalism the corresponding
leads to an alternative differential equation for anomalous diffusion. In Section (\ref{conc})
we  discuss our results.
\section{Fisher information measure as a lagrangian} \label{Fisher}
The Fisher information measure is defined through the integral
\begin{equation}
I=\int dx \left(\frac{dp(x)}{dx}\right)^2p^{-1}(x), \label{FI}
\end{equation}
where $p(x)$ denotes the probability distribution. In Ref. \cite{Vs} it was shown that
$I$ can be obtained  by considering a small global shift
$\Delta$ of $p(x)$ in the Kullback information as follows \footnote{An alternative derivation
is given in Ref. \cite{F1}}
\begin{equation}
I(\Delta)=\int dx  p(x)\ln[p(x)/p(x+\Delta)],
\end{equation}
and expanding in series in powers of the shift $\Delta$, leading to the equation
\begin{equation}
I(\Delta)=-\int dx \left[\frac{dp}{dx} \Delta -\left(\frac{(\frac{dp}{dx})^2}{p}-\frac{d^2p}{dx^2}\right)\frac{\Delta^2}{2}+...\right],
\end{equation}
where one can recognize the Fisher measure in the quadratic term.  In 1943, Fisher introduced \cite{Fisher}
in Equation (\ref{FI}) a new function $\phi(x)$, called probability amplitudes, in terms of $p(x)$
through the relation $p(x)=\phi^2(x)$ leading to the simpler  equation\footnote{Fisher used the symbol $q(x)$ instead of $\phi(x)$, which
we will not use to avoid confusion with the parameter $q$.}
\begin{equation}
I=4\int dx \frac{d\phi(x)}{dx}\frac{d\phi(x)}{dx},\label{phieq}
\end{equation}
which  is the starting point of
Frieden's approach \cite{F2} to derive the equations that describe the behavior of physical systems.
Introducing space-time coordinates, allowing two real components
$\phi_1$ and $\phi_2$, and after  defining $\Phi=\phi_1+i\phi_2$ Equation (\ref{phieq}) was promoted to a covariant
expression \cite{F3}
\begin{equation}
I=4\int d^4x \partial_{\nu}\Phi^{*} \partial^{\nu}\Phi,
\end{equation}
such that performing the Fourier transform of $I$ give us the energy constraint and thus
the mass term $m^2\Phi^{*}\Phi$, leading to the well known Klein-Gordon equation.
\section{Fisher information based on fractional calculus inspired entropy}\label{relat}
In a previous article \cite{Ubriaco}, based on the definition of the Riemann-Liouville 
fractional derivative, we proposed a new entropy function defined as
\begin{equation}
S_q[p]=\lim_{t\rightarrow -1}\frac{d}{dt}\left(_{-\infty}D_t^{q-1}\sum_ie^{-t\ln p_i}\right),\label{ent}
\end{equation}
where $0<q<1$, and the fractional derivative is given by
\begin{equation}
_aD_t^q=\left(\frac{d}{dt}\right)^n(_aD_t^{q-n}f(t)),
\end{equation}
with $n\in N$ and  $n>q$. The equation
\begin{equation}
_aD_t^{q-n}f(t)=\frac{1}{\Gamma(n-q)}\int_a^t\frac{f(t')}{(t-t')^{(1+q-n)}}dt'.\label{lhd}
\end{equation}
 defines the fractional integral operator.  Equation (\ref{ent}) leads to the entropy function
\begin{equation}
S_q[p]=\sum_i p_i (-\ln p_i)^q \label{ent2},
\end{equation}
which is a concave function and satisfies the Lesche and thermodynamic stability criteria.
As explained in \cite{Ubriaco}, the motivation to propose Equation (\ref{ent})
as a definition of a new entropy function is based on the observation that the Shannon entropy and Tsallis entropy
can be defined according to \cite{Abe}
\begin{eqnarray}
S&=&\lim_{t\rightarrow -1}\frac{d}{dt}\sum_i p_i^{-t}\nonumber\\
S_q &=&\lim_{t\rightarrow -1} D_q^t \sum_i p_i^{-t}, 
\end{eqnarray}
where the operator 
\begin{equation}
D_q^t=t^{-1}\frac{1-q^{td/dt}}{1-q}.
\end{equation} 
is called the Jackson q-derivative \cite{Jackson}.
With use of  Equation (\ref{ent2}) for two probability
distributions ${\bf p}=(p_1,...)$ and ${\bf P}=(P_1...)$ we define the corresponding relative entropy
as
\begin{equation}
H(p||P)=\sum_i p_i \left(-\ln\frac{P_i}{p_i}\right)^q.\label{relaten}
\end{equation}
The proof that $H(p||q)$ is positive follows along the same lines
as the $q=1$ case. Provided that $\sum_i p_i=1$ and $f(x)$ is a convex function, the Jensen inequality 
\begin{equation}
\sum_i p_i f(x_i)\geq f\left(\sum_i p_i x_i\right),
\end{equation}
and the fact that $f(x)=(-\ln x)^q$ is a convex function
allow us to write for Equation (\ref{relaten}),
\begin{eqnarray}
H(p||P)&\geq&\left(-\ln \sum_ip_i\frac{P_i}{p_i}\right)^q\nonumber \\
&=&(-\ln \sum_i P_i)^q\\
&=&0.\nonumber
\end{eqnarray}
Now, to extract the Fisher information measure for $q\neq 1$, we 
consider a distribution function $p(x)$ and a small shift $\Delta$ such that Equation (\ref{relaten}) becomes
\begin{equation}
H(p(x)||p(x+\Delta))=\int dx\; p(x) \left(-\ln\frac{p(x+\Delta)}{p(x)}\right)^q.
\end{equation}
Expanding $\ln(p(x+\Delta)$  up to second order in $\Delta$,
\begin{eqnarray}
H(p(x)||p(x+\Delta))&=&-\int dx \;p(x)\left[\frac{p'}{p} \Delta+\left(\frac{p''}{p}-\frac{p'^2}{p^2}\right)\frac{\Delta^2}{2}\right]^q \nonumber \\
&=&-\int dx\; p(x) \Delta^q \sum_{n=0}\left(_n^q\right)\left(\frac{p'}{p}\right)^{q-n}\left[\left(\frac{p''}{p}-\frac{p'^2}{p^2}\right) \frac{\Delta}{2}\right]^n ,\nonumber
\end{eqnarray}
where $p'=\frac{dp}{dx}$. In order to obtain the Fisher information it is enough to keep the two lowest order terms,
\begin{equation}
H(p(x)||p(x+\Delta))=-\int dx \left(\left(_0^q\right)\frac{p'^q}{p^{q-1}}\Delta^q-\left(_1^q\right)\left(\frac{p'^{q+1}}{p^q}
-\frac{p''p'^{q-1}}{p^{q-1}}\right)\frac{\Delta^{q+1}}{2}\right),\nonumber
\end{equation}
such that, up to a constant, we define
\begin{equation}
I_q=\int dx\; \frac{p'^{q+1}(x)}{p^q(x)}.\label{Iq}
\end{equation}
In our case, after introducing the new function $\phi(x)=p^\frac{1}{q+1}(x)$, Equation (\ref{Iq})
becomes
\begin{equation}
I_q=(q+1)^{q+1}\int dx \left(\frac{d\phi}{dx}\right)^{\frac{q+1}{2}}\left(\frac{d\phi}{dx}\right)^{\frac{q+1}{2}},\label{lag}
\end{equation}
which reduces to Equation (\ref{phieq}) when $q\rightarrow 1$. Setting $\delta I_q=0$ in Equation (\ref{lag}),
leads to the equation
\begin{equation} 
\left(\frac{d\phi(x)}{dx}\right)^{q-1}\frac{d^2\phi(x)}{dx^2}=0.\label{eqn}
\end{equation}
Although solutions to Equation (\ref{eqn}) are not different than the $q=1$ case, solutions
to the corresponding eigenvalue problem are different. Let us consider the equation
\begin{equation}
\left(\frac{d\phi(x)}{dx}\right)^{q-1}\frac{d^2\phi(x)}{dx^2}=\lambda \phi(x).\label{eigen}
\end{equation}
Solutions to Equation (\ref{eigen}) are given by the function
\begin{equation}
\phi(x)=A\left[1+(1-q)\gamma x\right]^{\omega},\label{sol}
\end{equation}
where $A$, $\gamma$ and $\omega$ are constants to be determined. 
Taking the elementary derivatives
we find that  $\omega=-\frac{q+1}{1-q}$ and $\gamma=\left(\frac{\lambda}{2(q+1)^{2q-1}}\right)^\frac{1}{2q}$,
where solutions with $\gamma>0$  restrict the values of $q$ such that $(-1)^\frac{q+1}{2q}=1$.
It is interesting to remark that functions  in Equation (\ref{sol}) are of the
type of solutions found for the probability distribution in the non-extensive formulation of statistical mechanics \cite{T1}\cite{T2}.
The constant $A$ is found by 
normalizing  the function in Equation (\ref{sol}) using  the definition of the $\Gamma$ function,
in a very similar way as done to calculate the partition function of non-extensive statistical mechanics
\cite{P}\cite{A},
\begin{equation}
\Gamma(z)=x^z\int_0^{\infty}t^{z-1}e^{-tx} dt\;\;\;x>0\;,\;z>0,
\end{equation}
 leading to the value $A=\gamma(1+q)$.

Taking the limit $q\rightarrow 1$ the function in Equation (\ref{sol}) becomes the
expected exponential solution $\phi(x;q=1)=\lambda^{1/2}exp(- \lambda^{1/2}x)$.

\section{Anomalous diffusion} \label{anomalous}

In the last twenty years, anomalous diffusion has been observed in a variety of interesting
physical systems, for example  breakable micelles dissolved in salted water \cite{OBLU},
measurements of cardiac beat to beat intervals in healthy individuals \cite{PMHHSG}, chaotic transport in a two-dimensional fluid flow in a rotating annulus \cite{SWS},  laser cooling of atoms to very low temperatures \cite{BBEAC}, and
more recently several interesting experiments with living cells in which a transition from sub-diffusive
to superdiffusive behavior is observed \cite{RZMMMF}\cite{GABR}.
On the other hand,
several mathematical models have been proposed
to study  different  aspects of anomalous diffusion.  Some of these models are based on
a linear differential equation for diffusion on fractals \cite{SP},  linear and non-linear fractional differential 
equations  \cite{S}-\cite{MK}, a study of particle chaotic dynamics along a stochastic web
 \cite{ZSW}, a study of super-diffusion in a hamiltonian system in the context of a continuous-time random 
walk approach to L\'{e}vy flights \cite{KZ}, nonlinear differential equations \cite{B}-\cite{SSLMM} and some calculations based on Tsallis
non-extensive statistical mechanics
\cite{PP}-\cite{S.Abe}.  

In our case, the simplest model we can propose is
based on the differential equation
\begin{equation}
K \left(\frac{\partial W}{\partial x}\right)^{q-1}\frac{\partial^2W}{\partial x^2}=\frac{\partial W}{\partial t},\label{anomeq}
\end{equation}
where $K$ is the diffusion constant. The solution  to the eigenvalue problem in the previous section
guides us to propose the function
\begin{equation}
W(x,t)=\frac{A}{t^{\lambda}}\Psi(x,t)^{\omega},\label{solan}
\end{equation}
where 
\begin{equation}
\Psi(x,t)=\left[1+\frac{(1-q)}{C}\frac{x^{\gamma}}{t^{\beta}}\right],
\end{equation}
and $A$, $\lambda$, $C$, $\beta$, $\gamma$ and $\omega$ are constants to be determined,
which in the $q\rightarrow 1$ limit must satisfy: $\lambda \rightarrow 1/2$, $\beta \rightarrow 1$ and
$\gamma \rightarrow 2$.  By performing the elementary derivatives, comparing powers in $\Psi(x,t)$,
$x$ and $t$ and the constants in both sides of Equation (\ref{anomeq}) we find that 
\begin{eqnarray}
\omega&=&-\frac{q}{1-q}\\
\gamma&=&\frac{q+1}{q}\\
\beta&=&\frac{q+1}{2q^2}\\
\lambda&=&\frac{1}{2q}\\
A^{1-q}&=&\frac{2K (q+1)^q}{C^q}.\label{ACeq}
\end{eqnarray}
Similarly to the previous section there is a restriction on the values of $q$,  due to the  fact that
in order to have the correct $q\rightarrow 1$ limit the constants $A$ and $C$ have to be positive leading to the
set of allowed values of $q=\frac{2n-1}{2n+1}$ where $n=1,2,3...$. Since the solution $W(x,t)$ must have
dimension $[W]=\frac{1}{cm}$, these constants have dimensions:
$[K]=\frac{cm^{2q}}{s}$, $[A]=\frac{s^{\lambda}}{cm}$ and $[C]=\frac{{cm}^{\gamma}}{s^{\beta}}$.
In Fig. 1 we show the function $W(x,t)$  for the case of $q=5/7$ and $t=0.05,0.09$ and $0.2$, and in Fig. 2
the function $W(x,t=0.5)$ for the superdiffusive cases $q=5/7$,$q=19/21$  and the  normal diffusive case $q=1$.

As a next step we find an additional relation between the constants $A$ and $C$ by normalizing the solution $W(x,t)$
\begin{equation}
\frac{A}{t^{\lambda}}\int_{-\infty}^{\infty} dx \left[1+\frac{(1-q)}{C}\frac{x^{\gamma}}{t^{\beta}}\right]^{\omega}=1.
\end{equation}
Since $q$ takes values such that $(-1)^{\gamma}=1$ we can change the limits from $(-\infty, \infty)$
to $(0, \infty)$, such that with use of the integral representation of the $\Gamma$ function we obtain
\begin{equation}
A=\left(\frac{1-q}{C}\right)^{1/\gamma}g(q),
\end{equation}
where the function $g(q)=\frac{\gamma \Gamma(-\omega)}{2\Gamma(1/\gamma)\Gamma(-\omega-1/\gamma)}$.
Defining $Q=\frac{2q^2}{1-q^2}$ and by using \cite{GR}
\begin{equation}
\lim_{Q\rightarrow \infty}\frac{\Gamma(Q+\frac{1}{\gamma})}{\Gamma(Q
)}e^{-(1/\gamma)\ln Q}=1,
\end{equation}
with Equation (\ref{ACeq}) we find that the constant $A$ has the correct
normal diffusion limit $A\rightarrow \sqrt{\frac{1}{4\pi K}}$ as $q\rightarrow 1$.
The constant $C$ is given by
\begin{equation}
C=(2K)^{\lambda\gamma}(1+q)^{\gamma/2}(1-q)^{1/2\omega}g^{\gamma/2\omega}(q).
\end{equation}
The mean squared displacement $<x^2>$ is obtained by solving
\begin{equation}
<x^2>= \frac{A}{t^{\lambda}}\int_{-\infty}^{\infty} dx \; x^2\left[1+\frac{(1-q)}{C}\frac{x^{\gamma}}{t^{\beta}}\right]^{\omega},
\end{equation}
leading, up to a $q$-dependent constant, to
\begin{equation}
<x^2>\sim K^{1/q}t^{1/q},\;\;\;\;\; 0<q<1.\label{sd}
\end{equation}
Since $0<q<1$,  our model seems to describe exclusively superdiffusive processes.
This apparent drawback is due to the fact that we took in Equation (\ref{ent}) the
particular value $n=1$.  A general expression, $\forall n>0$ for the entropy function $S_q[p]$
reads
\begin{equation}
S_q[p]=\lim_{t\rightarrow -1}\left(\frac{d}{dt}\right)^n\left(_{-\infty}D_t^{q-n}\sum_ie^{-t\ln p_i}\right),
\end{equation}
which leads to the same mathematical function
\begin{equation}
S_q[p]=\sum_i p_i (-\ln p_i)^q,
\end{equation}
but now the parameter $q$ takes values in the set $n-1<q<n$. 
The  solution for $q>1$ reads
\begin{eqnarray}
W'(x,t) &=& \frac{A'}{t^{\lambda}}\left[ 1-\frac{(q-1)}{C'}\frac{x^{\gamma}}{t^{\beta}}\right]^{\omega}\nonumber  \\
  &=&  0 \;\;\;\;\;\;\;\mbox{for $(1-\frac{(q-1)}{C'}\frac{x^{\gamma}}{t^{\beta}})<0$},\;\;\;\; q>1 ,\label{W'}
\end{eqnarray}
with the constants $A'$ and $C'$ given by
\begin{eqnarray}
A'&=&\left[\frac{\gamma\lambda}{\omega^qK(\gamma-1)}\right]^{\lambda}g'(q)^{\frac{\gamma}{2}}\nonumber\\
C'&=&\left[\frac{(\gamma-1)K}{\lambda}\right]^{\frac{1}{q}-\frac{\lambda}{\omega}}(q-1)\omega^{\frac{\gamma}{2}}\gamma^{1+\frac{\lambda}{\omega}}g'(q)^{\frac{\gamma}{2\omega}},
\end{eqnarray}
with $g'(q)=\frac{\Gamma(1+\omega+1/\gamma)}{2\Gamma(1/\gamma)\Gamma(1+\omega)}$.  In order to normalize
the new solution we  used the integral representation of the $\Gamma$-function in the complex plane
\begin{equation}
\frac{1}{\Gamma(z)}=\frac{i}{2\pi}\oint_C(-t)^{-z}e^{-t} dt.
\end{equation}
 In this case, $q>1$, the constant $\omega$ is positive and therefore the solution $W'(x,t)$ becomes the usual normal diffusion solution in the $q\rightarrow 1$ limit.  The allowed values of $q$ are $q=\frac{2n+1}{2n-1}$, and thus Equation (\ref{sd}) for an arbitrary positive integer $n$, includes also the case of subdiffusive behavior. Fig. 3 displays the function $W'(x,t=0.5)$ in Equation (\ref{W'}) for the subdiffusive cases $q=7/5$, $q=21/19$
 and the normal diffusive case $q=1$.

It is interesting to remark that for the particular value $\gamma=2$ and arbitrary values for the constants  $\omega$, $\lambda$ and $\beta$ the functions $W(x,t)$ and $W'(x,t)$ in Equations (\ref{solan})
and (\ref{W'})
 are of the same type of the solutions
of the nonlinear diffusion equation studied in \cite{B}\cite{CJ} \cite{V},  called the Barenblatt-Pattle solution.
\section{Conclusions}\label{conc}
In this paper,  based on a previously proposed 
entropy function $S_q[p]$ we first defined the corresponding relative entropy $H(p||P)$, and by 
considering the two probability distributions to differ by a small global shift $\Delta$,
we expanded the relative entropy with respect to this shift and extracted the Fisher information measure
for $q\neq 1$.
 Then, following Frieden's idea  we defined a new function $\phi(x)$ in terms
of the probability $p(x)$ that allowed us to reinterpret the Fisher information  as
 a free action $I_q$. Setting the variation of $I_q$ to $\delta I_q=0$  we find a differential operator that reduces to a
second order derivative at $q=1$. We found that the eigenfunctions of this differential operator are
of the same type than the probability distribution found in the Tsallis formulation of nonextensive statistical mechanics. Using this differential operator we proposed a simple differential equation that
coincides with  the normal diffusion equation in the $q\rightarrow 1$ limit. In our case, 
 the solutions are such the exponent $\gamma$ in the $x$ variable
is a fraction $\gamma=\frac{q+1}{q}$, taking
values in the set $\gamma=\frac{4n}{2n-1}$ for $q=\frac{2n-1}{2n+1}$ (superdiffusion), and $\gamma=\frac{4n}{2n+1}$ for $q=\frac{2n+1}{2n-1}$ (subdiffusion),$\forall n=1,2,3...$. The work done in this paper
is a first step to study anomalous diffusion with a rather simple and different mathematical model. In order to
study anomalous diffusion in a more realistic way will require to introduce a time dependent source term
(absorption) and external forces (drift) in a three dimensional version of this model, which we will attempt
to address in future communications.


\begin{thebibliography}{99}
\bibitem{T2} A list of works can be found in tsallis.cat.cbpf.br/biblio.htm.
\end{thebibliography}
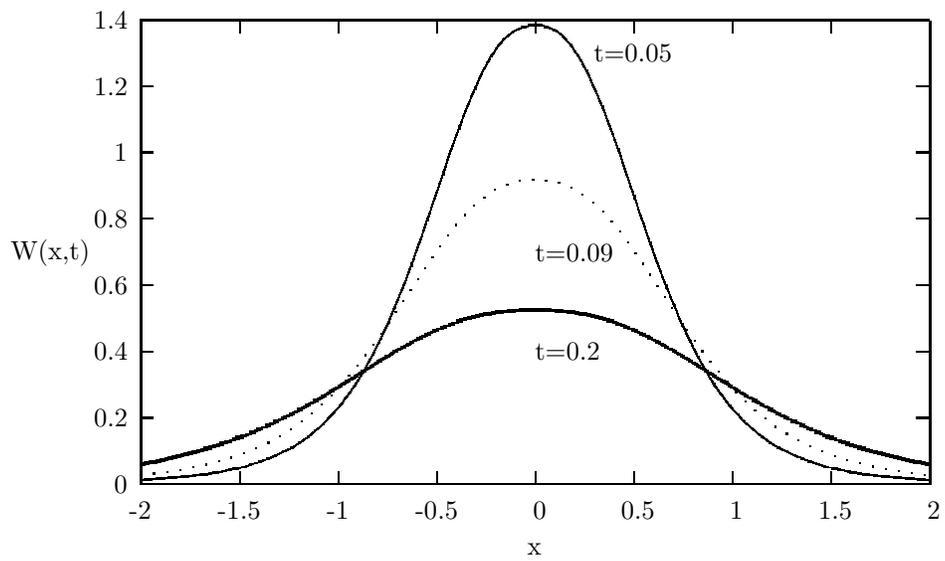
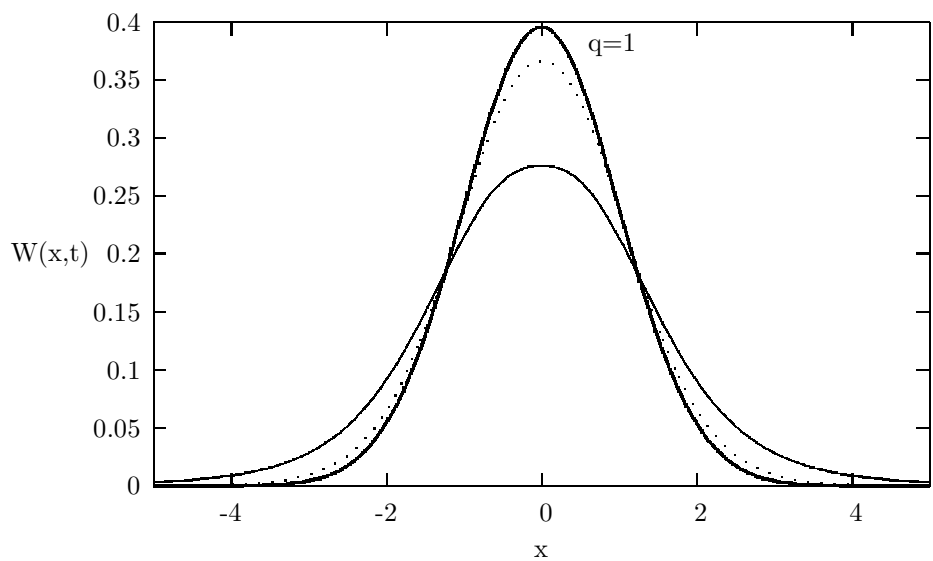
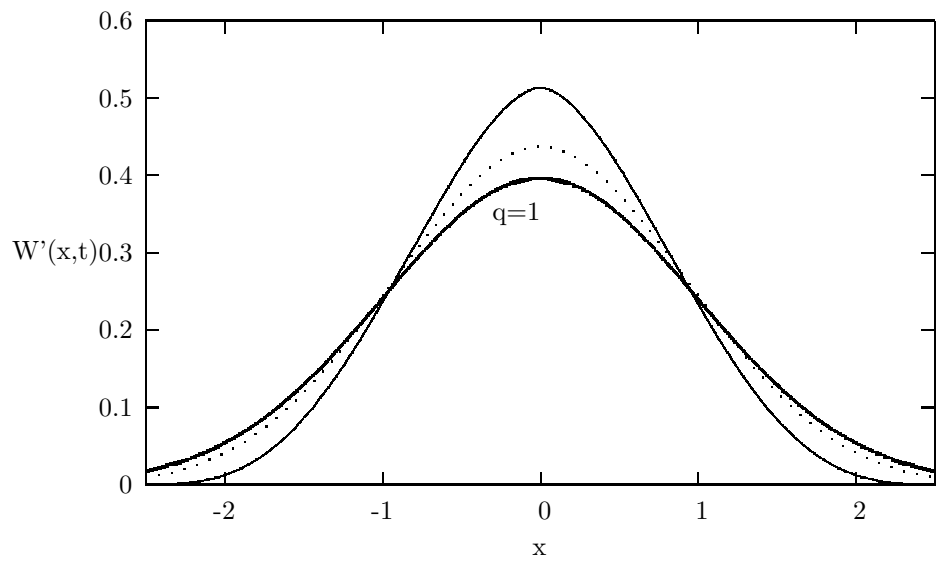
\end{document}